\documentclass[aps,prl,twocolumn,showpacs,superscriptaddress,groupedaddress]{revtex4}  
\usepackage{graphicx}  
\usepackage{dcolumn}   
\usepackage{bm}        
\usepackage{amssymb}   
\usepackage{subfigure}
\usepackage{color}
\usepackage{epstopdf}
\usepackage{threeparttable}
\usepackage{indentfirst}
\usepackage{soul}
\usepackage{tikz}
\usepackage{listings}
\usepackage{multirow}
\usepackage{amsmath}
\usepackage{float}
\usepackage{amsfonts}
\tikzset{>=stealth}

\begin{document}

\widetext
\leftline{Version xx as of \today}
\leftline{To be submitted to PRL.)}


\title{The Influences of Edge Asymmetry on Network Robustness}
\author{Lei Wang}
\affiliation{Beihang University, China}
\author{Xinchen Wang}
\affiliation{Beihang University, China}
\date{\today}

\begin{abstract}
Asymmetry of in/out-degree distribution is a widespread phenomenon in real-world complex networks. This paper put forward the concept of Edge Asymmetry(EA) to quantify this feature. We designed an EA-based strategy to attack six kinds of real-world networks and found that it was able to achieve the effect as well as edge betweenness-based(EB) and better than edge degree-based(ED) and random attack strategies. In simulation, we found that the greater the network asymmetry the better the EA-based attack strategy performed. By definition, the computational complexity of EA was much lower than that of EB. Therefore, EA-based attack strategies were superior in efficiency. We verified the effect of the EA-based attack strategy with four groups of large-scale networks.
\end{abstract}

\maketitle


The robustness of a complex network indicates its ability to resist attacks. Many have studied the robustness property by evaluating the network connectivity after removing some nodes or edges \cite{Albert2000Error}\cite{crucitti2004error}\cite{Shargel2003Optimization}. We focus on the influences of edge attack strategies. Strategies based on random, degree and betweenness have been studied by attacking various complex network models and real-world networks \cite{Holme2002Attack}\cite{kurant2007error}\cite{iyer2013attack}.  There are, however, two problems. First, the random and ED strategies are fast but their effects are relatively poor. Second, the EB strategy achieves relatively better attack effect, but it is high in computational complexity and not applicable to large-scale networks. In this paper, we analyze the network structures and propose a novel attack strategy with better effect and high efficiency.



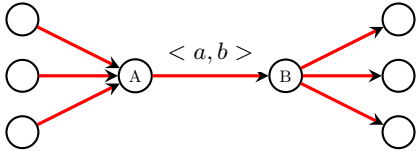
\begin{figure}[htbp]
\begin{center}

\begin{tikzpicture}
\begin{scope}[every node/.style={circle,thick,draw,,scale=.7}]
    \node (A) at (0,0) {A};
    \node (B) at (2,0) {B};
    \node[text height=10pt] (C) at (-1.5,0.75) {};
    \node[text height=10pt] (D) at (-1.5,0) {};
    \node[text height=10pt] (E) at (-1.5,-0.75) {};
    \node[text height=10pt] (F) at (3.5,.75) {} ;
    \node[text height=10pt] (G) at (3.5,0) {} ;
    \node[text height=10pt] (H) at (3.5,-.75) {} ;
\end{scope}

\begin{scope}[every node/.style={fill=white},
              every edge/.style={draw=red,very thick}]
    \path [->] (A) edge node[above = 1pt, text height=2pt] {$<a, b>$} (B);
    \path [->] (C) edge (A);
    \path [->] (D) edge (A);
    \path [->] (E) edge (A);
    \path [->] (B) edge (F);
    \path [->] (B) edge (G);
    \path [->] (B) edge (H);
\end{scope}
\end{tikzpicture}
\caption{A typical asymmetric network structures. Node a's in-degree is larger than its out-degree. In Node b, it is the opposite.}\label{fig:1}
\end{center}
\end{figure}
\vspace{-0.7cm}

The distribution of out-degree differ greatly from that of in-degree in many real-world networks, such as industrial networks \cite{Luo2017Asymmetry}, social networks \cite{Wei2016Handling}, WWW \cite{albert2002statistical} and software networks \cite{myers2003software,Wang2014}. In an asymmetric network, such as one shown in Figure \ref{fig:1}, we observed intuitively that removing the edge of $<a, b>$ would cause greater damage to the network structure than removing one of the other edges. Inspired by this observation, we proposed measuring the asymmetry of an edge with Edge Asymmetry(EA). EA and the nodes connected to the edge were closely related, so we first defined Node Asymmetry (NA).

\setlength{\abovedisplayskip}{1pt}
\begin{eqnarray}
{NA}_{a}^{in} = \frac{k_{a}^{in}}{D_{a}}\\
{NA}_{a}^{out} = \frac{k_{a}^{out}}{D_{a}}
\end{eqnarray}
\setlength{\belowdisplayskip}{1pt}

Here $k_{in}^{a}$ and $k_{out}^{a}$ are the in-degree and out-degree of Node a, $D_{a}$ is the degree of Node a.

As shown in Figure \ref{fig:1}, EA of the edge $<a, b>$ can be divided into in-degree $EA^{in}$ and out-degree $EA^{out}$ according to node's in-degree and out-degree.

\setlength{\abovedisplayskip}{1pt}
\begin{eqnarray}
EA_{ab}^{in}=NA_{a}^{in}- NA_{b}^{in}\\
EA_{ba}^{out}=NA_{b}^{out}- NA_{a}^{out}
\end{eqnarray}
\setlength{\belowdisplayskip}{1pt}

Noted that in-degree $EA^{in}$ and out-degree $EA^{out}$ have the same significance mathematically.

\setlength{\abovedisplayskip}{1pt}
\begin{eqnarray}
\begin{aligned}
EA_{ab}^{in}&=NA_{a}^{in}- NA_{b}^{in}\\
&=\frac{k_{a}^{in}}{D_{a}}-\frac{k_{b}^{in}}{D_{b}}\\
&=\frac{k_{a}^{in}D_{b}-k_{b}^{in}D_{a}}{D_{a}D_{b}}\\
\end{aligned}
\end{eqnarray}
\setlength{\belowdisplayskip}{1pt}

\setlength{\abovedisplayskip}{1pt}
\begin{eqnarray}
\begin{aligned}
EA_{ba}^{out}&=NA_{b}^{out}- NA_{a}^{out}\\
&=\frac{k_{b}^{out}}{D_{b}}-\frac{k_{a}^{out}}{D_{a}}\\
&=\frac{k_{b}^{out}D_{a}-k_{a}^{out}D_{b}}{D_{a}D_{b}}\\
&=\frac{(D_{b}-k_{b}^{in})D_{a}-(D_{a}-k_{a}^{in})D_{b}}{D_{a}D_{b}}\\
&=\frac{D_{a}D_{b}-k_{b}^{in}D_{a}-D_{a}D_{b}+k_{a}^{in}D_{b}}{D_{a}D_{b}}\\
&=\frac{k_{a}^{in}D_{b}-k_{b}^{in}D_{a}}{D_{a}D_{b}}\\
\end{aligned}
\end{eqnarray}
\setlength{\belowdisplayskip}{1pt}

As the in-degree $EA^{in}$  and out-degree $EA^{out}$ of the edge $<a, b>$ in Figure \ref{fig:1} have the same value, we selected in-degree $EA^{in}$ to represent EA.


Complex networks exhibit different levels of robustness under the various attacks \cite{Albert2000Error}\cite{crucitti2004error}. The concept of network asymmetry inspired us to try to
attack by EA. We compare it with three other strategies with same dataset. The details of these strategies are described as follows.

(a) Edge Asymmetry (EA): Edge asymmetry reflects the significance of edge in the network connectivity. The edge with the largest asymmetry will be deleted in each attack.

(b) Edge Degree (ED): Degree is a critical attribute of a node \cite{Strogatz2000Network}. The degree-based attack strategy is the most widely-used attack method. It is proposed by Ref. \cite{Holme2002Attack} where the edge degree is defined by Eq.(\ref{e.ed}).

\setlength{\abovedisplayskip}{2pt}
\begin{eqnarray}
 k^e_{ed} = k^v_d \times k^w_d
\label{e.ed}
\end{eqnarray}
\setlength{\belowdisplayskip}{2pt}

where the edge \emph{e} connects node \emph{v} and \emph{w}, $k^e_{ed}$ denotes the ED of edge $e$, and $k^v_d$ and $k^w_d$ represent the degree of the node $v$ and $w$, respectively. The strategy selects the edges in descending degrees.

(b) Edge Betweenness (EB): Edge betweenness of the edge $<a, b>$ in the graph  G=$<V,E>$  is defined as:

\setlength{\abovedisplayskip}{2pt}
\begin{eqnarray}
B_{ab}=\frac{\sum_{n}^{g}\sum_{n}^{k}\frac{L(g,k,e_{ab}))}{L(g,k))}}{n(n-1))/2}
\label{e.eb}
\end{eqnarray}
\setlength{\belowdisplayskip}{2pt}

In Eq.(\ref{e.eb}), \emph{L(g,k)} represents the number of shortest paths between node \emph{g} and \emph{k}, $L(g,k,e_{ab})$ represents the number of paths passing edge $e_{ab}$, and \emph{n} is the total number of nodes in the graph. In most real-world large-scale networks, the importance of each edge is different. The betweenness centrality of an edge is its general static characteristic quantity, reflecting the status and influence of the edge in the whole network \cite{He2009Betweenness}. In this strategy, the edge with the largest betweenness will be deleted in each attack.

(d) Random Attack (RA): Random attack takes no account of edge attributes, and deletes the randomly selected edge in each attack.

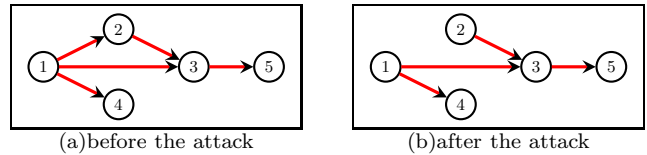
\begin{figure}[htbp]
\begin{center}
\subfigure[before the attack]{
\fbox{
\begin{tikzpicture}[scale=.5]
\begin{scope}[every node/.style={circle,thick,draw, scale=.7}]
    \node (1) at (0,0) {1};
    \node (2) at (2,1) {2};
    \node (3) at (4,0) {3};
    \node (4) at (2,-1) {4};
    \node (5) at (6,0) {5};
\end{scope}

\begin{scope}[every node/.style={fill=white},
              every edge/.style={draw=red,very thick}]
    \path [->] (1) edge (2);
    \path [->] (1) edge (4);
    \path [->] (1) edge (3);
    \path [->] (2) edge (3);
    \path [->] (3) edge (5);
\end{scope}
\end{tikzpicture}
}
}
\hfill
\subfigure[after the attack]{
\fbox{
\begin{tikzpicture}[scale=.5]
\begin{scope}[every node/.style={circle,thick,draw, scale=.7}]
    \node (1) at (0,0) {1};
    \node (2) at (2,1) {2};
    \node (3) at (4,0) {3};
    \node (4) at (2,-1) {4};
    \node (5) at (6,0) {5};
\end{scope}

\begin{scope}[every node/.style={fill=white},
              every edge/.style={draw=red,very thick}]
    \path [->] (1) edge (4);
    \path [->] (1) edge (3);
    \path [->] (2) edge (3);
    \path [->] (3) edge (5);
\end{scope}
\end{tikzpicture}
}
}
\caption{An Example of Network Attack. As shown in Figure \ref{fig:2}(a), before the attack, the largest weakly connected subgraph and the largest spanning tree each consists of five nodes, and the largest strongly connected subgraph has one node. After removing the edge $<1,2>$ , as shown in Figure \ref{fig:2}(b), the largest weakly connected subgraph in the network still holds five nodes, the largest strongly connected subgraph still has one node, but the largest spanning tree has four nodes. } \label{fig:2}
\end{center}
\end{figure}

\begin{figure*}[htbp]
\centering
\subfigure[linux]{\includegraphics[width=.27\textwidth]{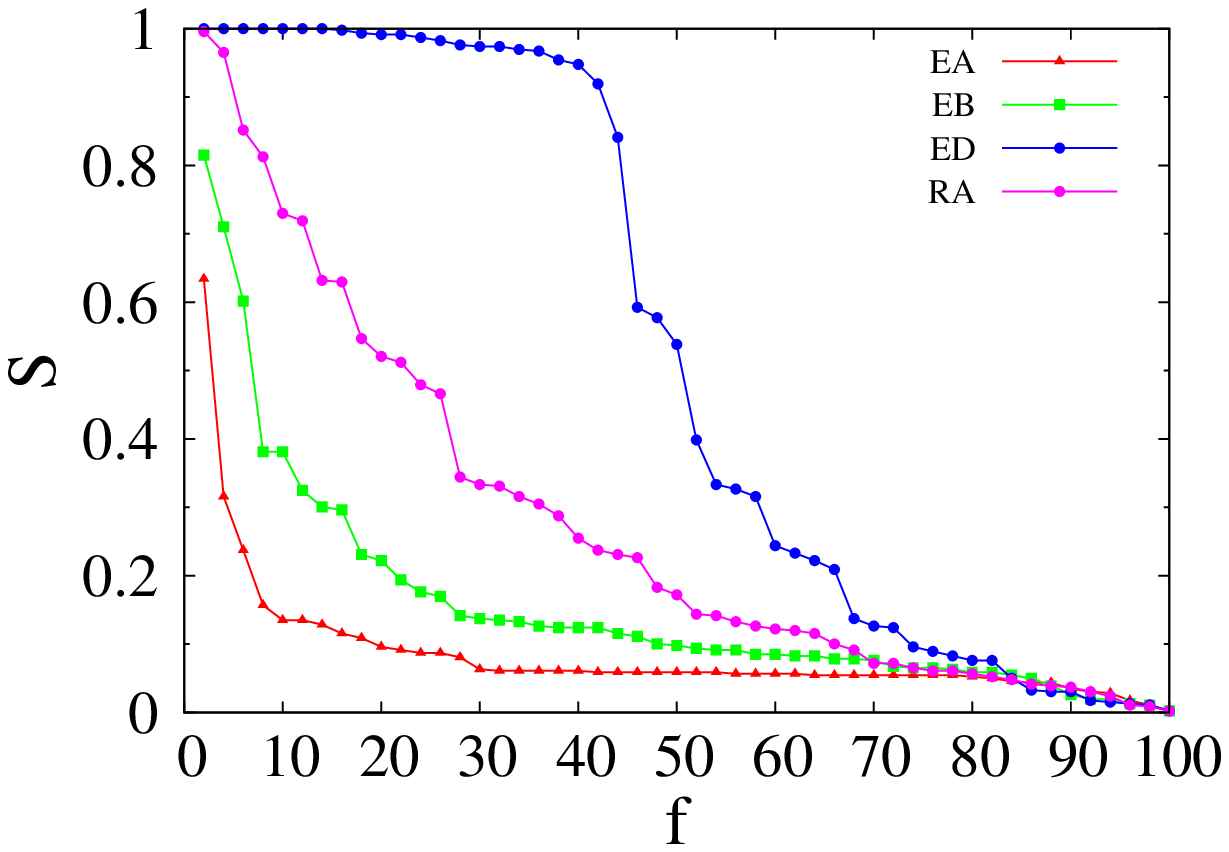}}
\subfigure[cit-hepth]{\includegraphics[width=.27\textwidth]{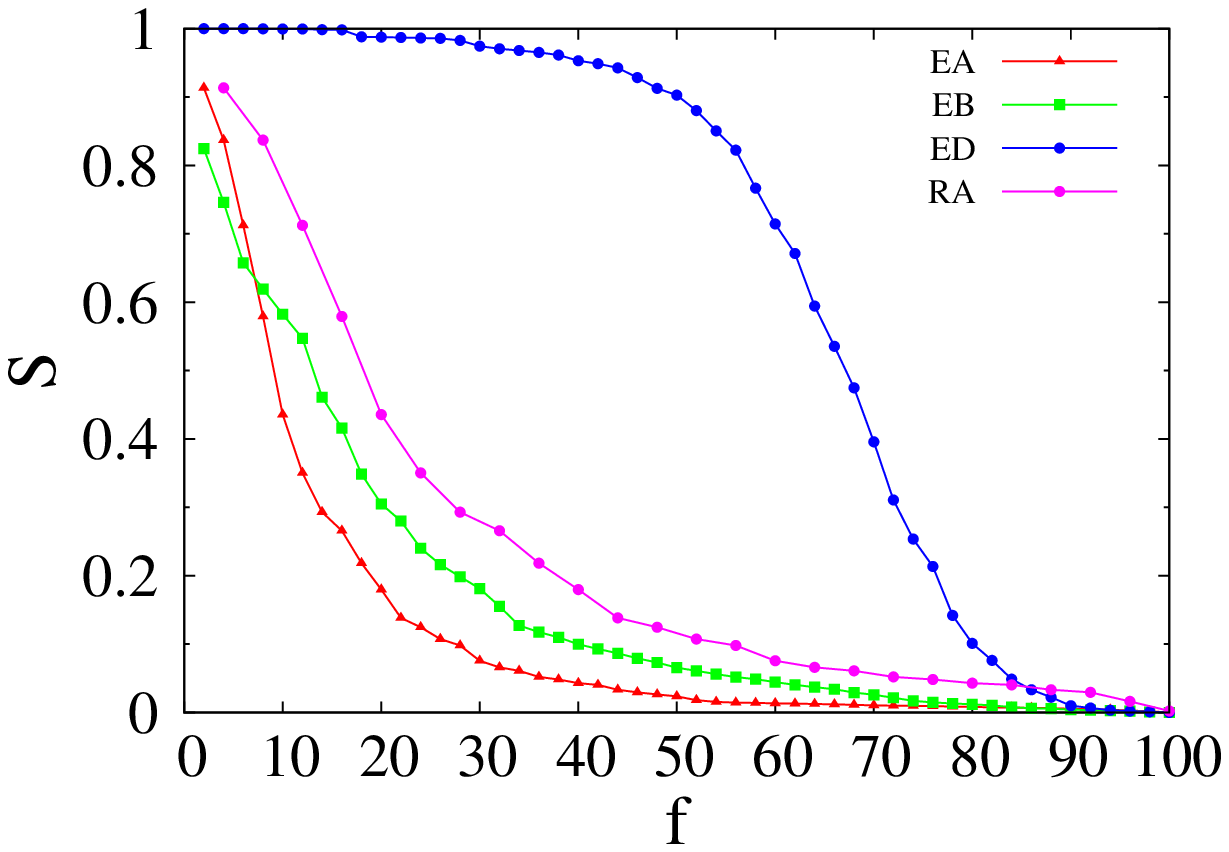}}
\subfigure[LittleRockLake]{\includegraphics[width=.27\textwidth]{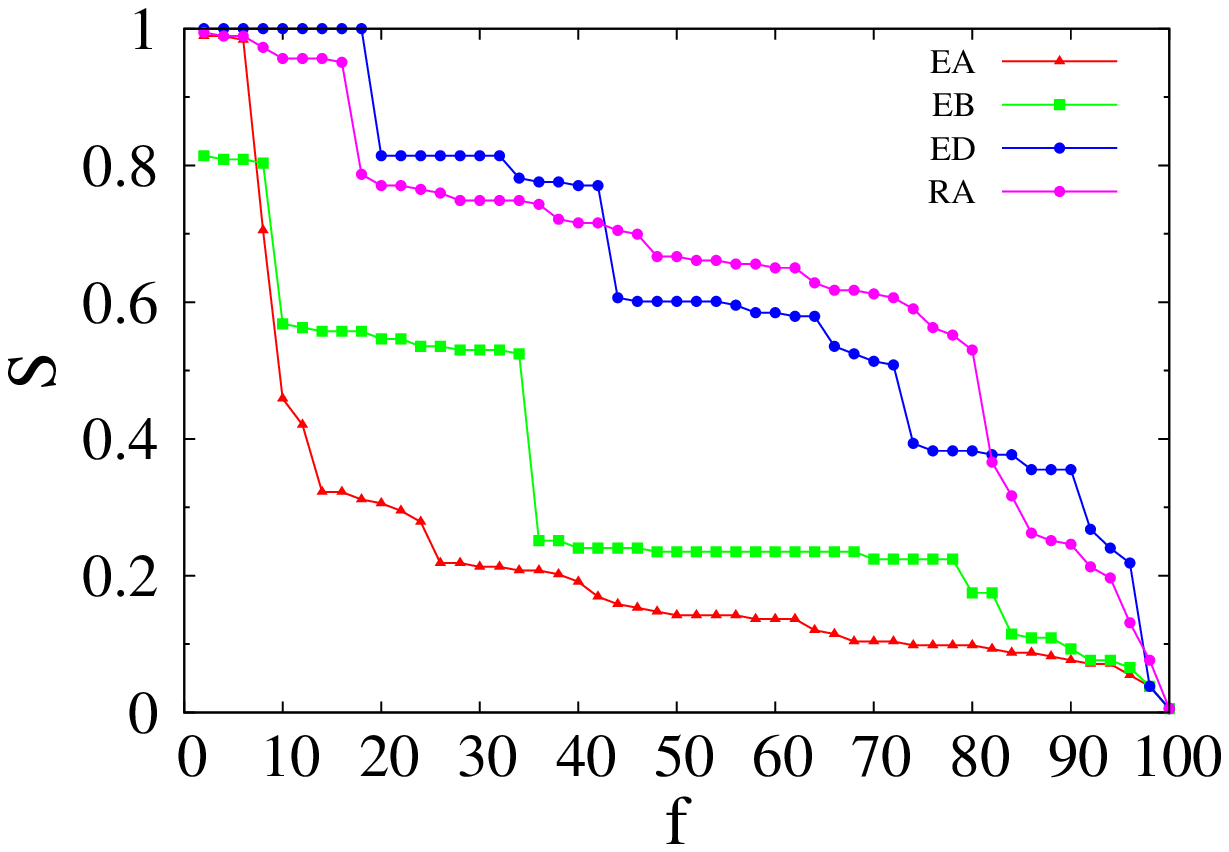}}
\subfigure[celegansneural]{\includegraphics[width=.27\textwidth]{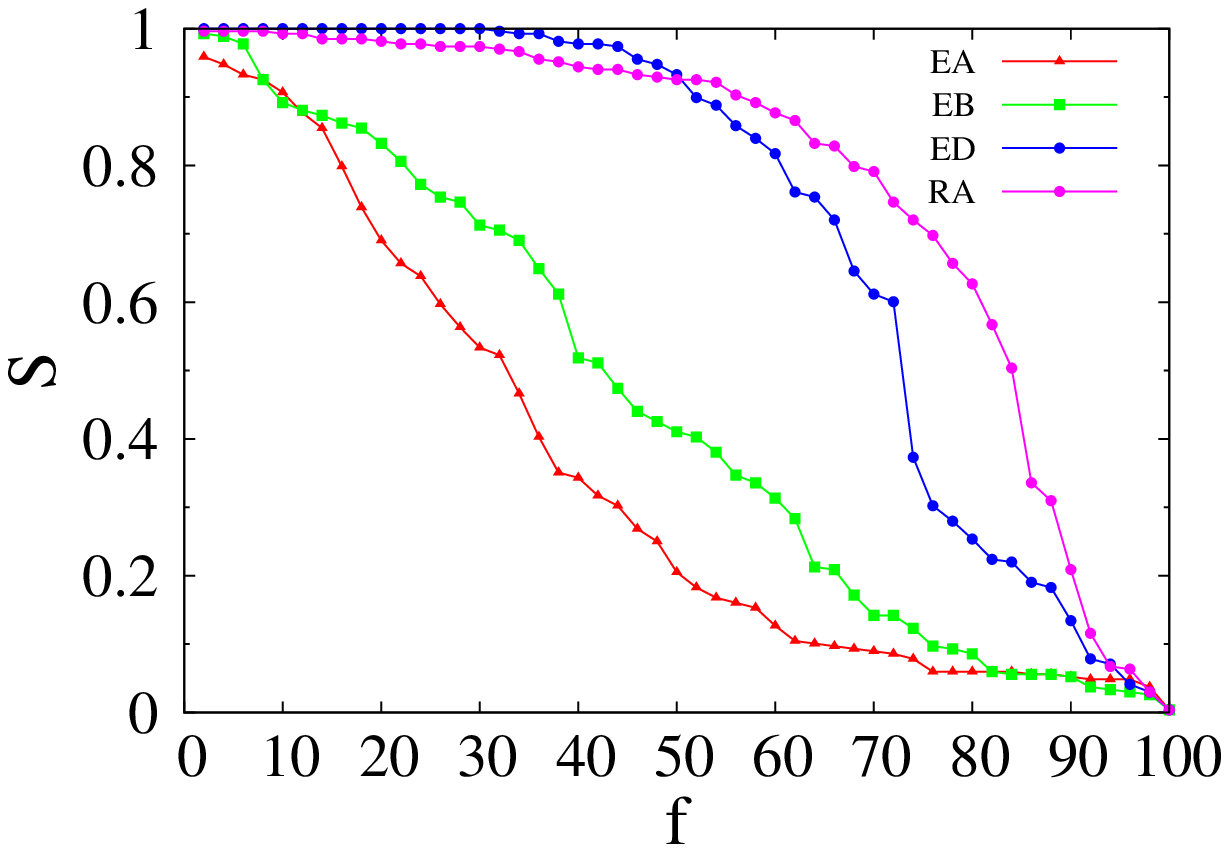}}
\subfigure[polblogs]{\includegraphics[width=.27\textwidth]{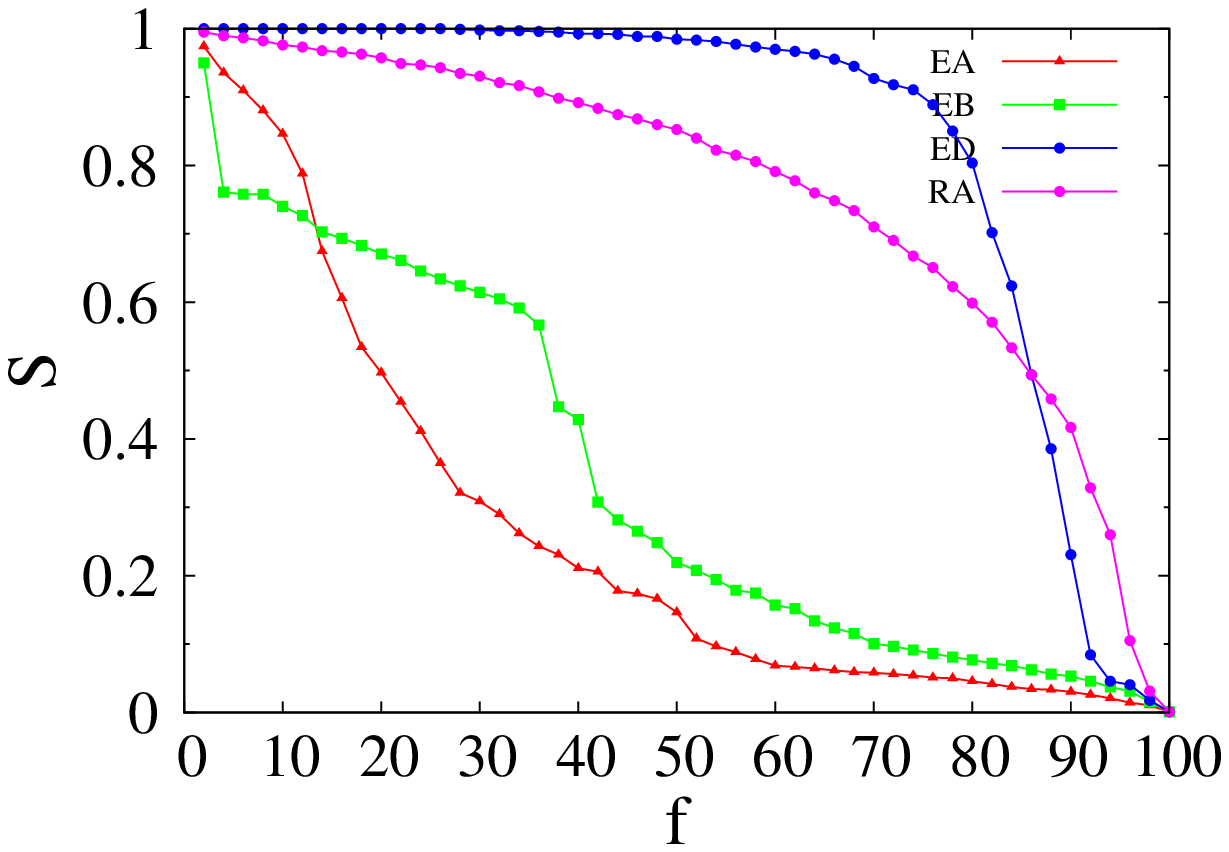}}
\subfigure[macaque]{\includegraphics[width=.27\textwidth]{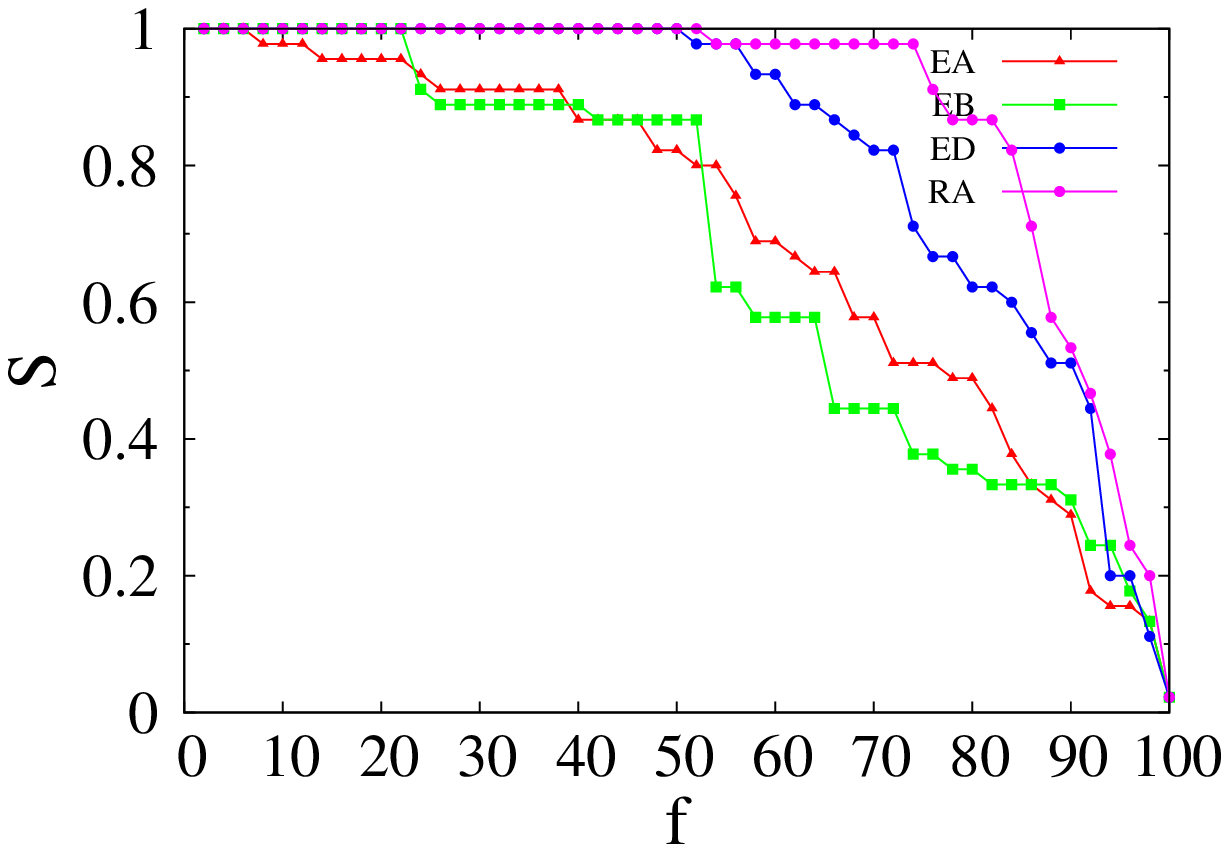}}
\caption{Results of real network attack experiment. The abscissa f represents the percentage of the removed edges in the networks, and the ordinate S the attack effect. The data sets are Linux l.2.0 function call graph\cite{Wang2014} with 3992 nodes and 14954 edges, cit-Hepth\cite{Leskovec2005Graphs} with 34,546 nodes and 421,578 edges, LittleRockLake\cite{Martinez1991Artifacts}  with 183 nodes and 2494 edges, polblogs\cite{Adamic2005The} with 297 nodes and 2345 edges, celegansneural\cite{Watts1998Collective} with 1224 nodes and 19025 edges and macaque\cite{Patric2008Mapping} with 45 nodes and 463 edges.} \label{fig:3}
\end{figure*}

We quantified the robustness of large networks by studying how well these networks stay connected under edge-removal attacks. Generally, the scale of strongly connected subgraph, weakly connected subgraph and spanning tree \cite{Camerini1978The} are used to evaluate the robustness of networks. For directed graphs, we found that we could better measure the attack effect using the largest spanning tree.


When removing a few edges to change the network structure, the largest weakly connected subgraph is very likely to remain unchanged because that weakly connected graph is nondirectional. Since strongly connected sub-graph can only include a small portion of the nodes in the directed graph because its edges are directional, it cannot represent the connectivity of the whole network. The largest spanning tree, however, is sensitive to the decrease of the edges in a network. Figure \ref{fig:2} shows an example where compared with the largest strongly connected subgraph and the largest weakly connected subgraph, the largest spanning tree can better reflect the change of the directed network's structure. Therefore, we chose the scale of the largest
spanning tree to measure the attack effect.

We use $S_i$ to represent the number of nodes in the maximum spanning tree after \emph{i}'th edge-removal attack, and $S_0$ denotes the number of nodes in the maximum spanning tree in the initial network. Then the ratio to quantify the robustness of networks, $S$, can be defined using Eq. (\ref{e.s}):

\setlength{\abovedisplayskip}{2pt}
\begin{eqnarray}
S=\frac{S_{i}}{S_{0}}
\label{e.s}
\end{eqnarray}
\setlength{\belowdisplayskip}{2pt}

When the same number of edges are removed from the network, the smaller the value of \emph{S}, the more vulnerable the network becomes, and the more efficient is the attack strategy.


Except random attack strategy, for each of attack strategies based on edge degree, edge betweenness and edge asymmetry, we first sorted the edges according to the edge attributes from the greatest to the smallest in the initial network. Next we deleted the top 2\% of the sorted edges for each round of attack. After the \emph{i}'th attack, we calculated the scale of the largest spanning tree $S_{i}$ in the network and compared $S_{i}$ with the largest spanning tree $S_{0}$ in the initial network to derive $S_{i}/S_{0}$.

Figure \ref{fig:3} showed the result of using four attack strategies to attack some real data sets. We could see that attack strategies based on EA and EB were more efficient for all data sets. EA attacks on networks such as Linux, cit-Hepth, polblogs and celegansneural seemed more effective than those using the EB strategy. Initially, EA strategy in LittleRockLake network seemed inferior to EB, but it quickly led to the collapse of the whole network. As for macaque, the EA strategy seemed inferior to the EB strategy. macaque has only 45 nodes, however, and  the attack effects of EA and EB strategies were hard to distinguish. Figure \ref{fig:3} also showed that the ED and RA strategies were less efficient on all data sets.


Next, we used some network examples to analyze the results of the four attack strategies to explain why the EA strategy was more effective.

The largest spanning tree of Figure \ref{fig:4}(b) has 6 nodes, and that of Figure \ref{fig:4}(c) has only 4. This indicates that the EA is superior to the EB attack strategy in this case. Edge with the highest betweenness is the target of each edge betweenness-based attack strategy. For a network formed by starting and diverging from a few nodes, the edge with the highest betweenness is generally the initial ones. The effect of removing this edge on the network structure is very small. Therefore, we think that the EB attack strategy assigns too much importance to the significance of the edges, and neglected its role in network connectivity. The EA attack strategy can better identify the edge most important to network connectivity.

\begin{figure}[htbp]
\begin{center}
\subfigure[before attack]{
\fbox{
\begin{tikzpicture}[scale=.5]
\begin{scope}[every node/.style={circle,thick,draw,scale=.7}]
    \node (1) at (0,0) {1};
    \node (2) at (2,0) {2};
    \node (3) at (4,0) {3};
    \node (4) at (4,1.5) {4};
    \node (5) at (4,-1.5) {5};
    \node (6) at (6,1.5) {6};
    \node (7) at (6,-1.5) {7};
\end{scope}

\begin{scope}[every node/.style={fill=white},
              every edge/.style={draw=red,very thick}]
    \path [->] (1) edge (2);
    \path [->] (2) edge (3);
    \path [->] (2) edge (4);
    \path [->] (2) edge (5);
    \path [->] (3) edge (6);
    \path [->] (3) edge (7);
\end{scope}
\end{tikzpicture}
}
}
\hfill
\subfigure[after EB attack]{
\fbox{
\begin{tikzpicture}[scale=.5]
\begin{scope}[every node/.style={circle,thick,draw, scale=.7}]
    \node (1) at (0,0) {1};
    \node (2) at (2,0) {2};
    \node (3) at (4,0) {3};
    \node (4) at (4,1.5) {4};
    \node (5) at (4,-1.5) {5};
    \node (6) at (6,1.5) {6};
    \node (7) at (6,-1.5) {7};
\end{scope}

\begin{scope}[every node/.style={fill=white},
              every edge/.style={draw=red,very thick}]
    \path [->] (2) edge (3);
    \path [->] (2) edge (4);
    \path [->] (2) edge (5);
    \path [->] (3) edge (6);
    \path [->] (3) edge (7);
\end{scope}
\end{tikzpicture}
}
}
\subfigure[after EA attack]{
\fbox{
\begin{tikzpicture}[scale=.5]
\begin{scope}[every node/.style={circle,thick,draw, scale=.7}]
    \node (1) at (0,0) {1};
    \node (2) at (2,0) {2};
    \node (3) at (4,0) {3};
    \node (4) at (4,1.5) {4};
    \node (5) at (4,-1.5) {5};
    \node (6) at (6,1.5) {6};
    \node (7) at (6,-1.5) {7};
\end{scope}

\begin{scope}[every node/.style={fill=white},
              every edge/.style={draw=red,very thick}]
    \path [->] (1) edge (2);
    \path [->] (2) edge (4);
    \path [->] (2) edge (5);
    \path [->] (3) edge (6);
    \path [->] (3) edge (7);
\end{scope}
\end{tikzpicture}
}
}
\caption{Comparison of the EA and EB attack strategies. (a) shows the original network. (b) shows the network structure after removing the edge of the largest edge betweenness, $<1, 2>$. (c) shows the network structure after the removal of the edge of the largest EA $<2, 3>$.} \label{fig:4}
\end{center}
\end{figure}
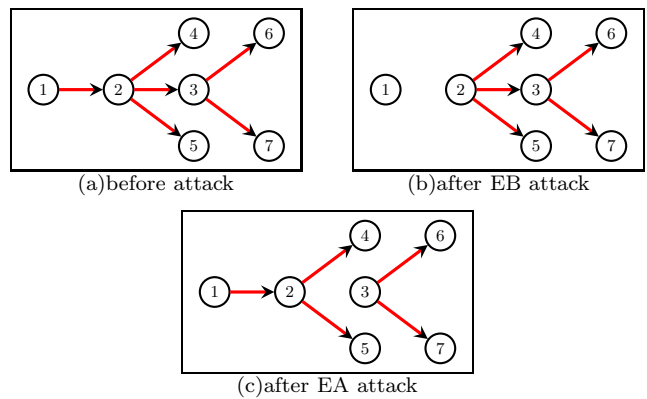


\begin{figure}[htbp]
\begin{center}
\subfigure[before attack]{
\fbox{
\begin{tikzpicture}[scale=.5]
\begin{scope}[every node/.style={circle,thick,draw, scale=.7}]
    \node (1) at (0,0) {1};
    \node (2) at (1.5,0) {2};
    \node (3) at (2.25,1.5) {3};
    \node (4) at (3,0) {4};
    \node (5) at (4.5,0) {5};
    \node (6) at (6,0) {6};
\end{scope}

\begin{scope}[every node/.style={fill=white},
              every edge/.style={draw=red,very thick}]
    \path [->] (1) edge (2);
    \path [->] (2) edge (3);
    \path [->] (3) edge (4);
    \path [->] (2) edge (4);
    \path [->] (4) edge (5);
    \path [->] (5) edge (6);
\end{scope}
\end{tikzpicture}
}
}
\hfill
\subfigure[after ED attack]{
\fbox{
\begin{tikzpicture}[scale=.5]
\begin{scope}[every node/.style={circle,thick,draw, scale=.7}]
    \node (1) at (0,0) {1};
    \node (2) at (1.5,0) {2};
    \node (3) at (2.25,1.5) {3};
    \node (4) at (3,0) {4};
    \node (5) at (4.5,0) {5};
    \node (6) at (6,0) {6};
\end{scope}

\begin{scope}[every node/.style={fill=white},
              every edge/.style={draw=red,very thick}]
    \path [->] (1) edge (2);
    \path [->] (2) edge (3);
    \path [->] (3) edge (4);
    \path [->] (4) edge (5);
    \path [->] (5) edge (6);
\end{scope}
\end{tikzpicture}
}
}
\subfigure[after EA attack]{
\fbox{
\begin{tikzpicture}[scale=.5]
\begin{scope}[every node/.style={circle,thick,draw,scale=.7}]
    \node (1) at (0,0) {1};
    \node (2) at (1.5,0) {2};
    \node (3) at (2.25,1.5) {3};
    \node (4) at (3,0) {4};
    \node (5) at (4.5,0) {5};
    \node (6) at (6,0) {6};
\end{scope}

\begin{scope}[every node/.style={fill=white},
              every edge/.style={draw=red,very thick}]
    \path [->] (1) edge (2);
    \path [->] (2) edge (3);
    \path [->] (3) edge (4);
    \path [->] (2) edge (4);
    \path [->] (5) edge (6);
\end{scope}
\end{tikzpicture}
}
}
\caption{Comparison of the EA and ED attack strategies. (a) shows the original network. (b) shows the network structure after removing the edge $<2, 4>$ of the largest degree. (c) shows the network structure after removing the edge $<2, 3>$ of the largest EA.} \label{fig:5}
\end{center}
\end{figure}
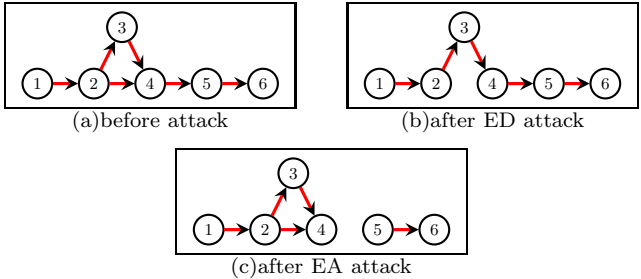

The largest spanning tree of Figure \ref{fig:5}(b) has 6 nodes, and that of \ref{fig:5}(c) has only 4. This indicates that EA attack strategy is superior to the ED attack strategy. As for directed graph, merely considering degree can fail to correctly measure the importance of an edge in network connectivity. For example, the edge $<2, 4>$ in Figure \ref{fig:5}(a) has the largest degrees, but there are relatively more edges between node 2 and 4 to keep them connect. Therefore, removing such edges from the network will not cause great impact on network connectivity. EA is better able to recognize such edges and will not mistakenly attack them.


Figure \ref{fig:3} showed that the EA attack strategy was effective on real-world networks, which preliminarily confirmed our speculation, i.e., network asymmetry had a non-ignorable influence on network vulnerability. Next, we set out to verify whether the effect of network attack was better when the asymmetry is higher. We designed a graph generation algorithm, which controlled the asymmetry of generated graph, and then attacked the networks of different asymmetry using the EA strategy. We used the random graph generation algorithm in Ref. \cite{Newman2001Random}. Additionally, we adjusted the network asymmetry by controlling the relative size of out-degree and in-degree ranges of nodes. The graph generation algorithm is described as follows:

\begin{enumerate}
  \item Create an empty in-degree queue $d_{in}$ and an empty out-degree $d_{out}$.
  \item Randomly select a number $in_{i}$  from the in-degree range $[in_{min}, in_{max}]$ as the in-degree of the \emph{i}'th node and add it to $d_{in}$. Repeat until certain number of nodes are added (we set the number to around 800).
  \item Randomly select a number $out_{i}$ from the out-degree range $[out_{min}, out_{max}]$ as the out-degree of the \emph{i}'th node and add it to $d_{out}$. Repeat until the sum of $d_{out}$ is equal to that of $d_{in}$.
  \item Transform $d_{in}$ and $d_{out}$ into the representation of nodes. For example, transforming $d_{in}$   [2,3,4,2] into $n_{in}$ [0,0,1,1,1,2,2,2,2,3,3] means that No. 0 node has two in-degrees, No. 1 node has three in-degrees and No. 3 node has two in-degrees. Transforming $d_{out}$ [1,3,2,5] into $n_{out}$ [0,1,1,1,2,2,3,3,3,3,3] means that No. 0 node has one out-degree, No. 1 node has three out-degrees, No. 2 node has two out-degrees and No. 3 node has five out-degrees. In this case, node queues $n_{in}$ and $n_{out}$ are derived. In the end, their orders are randomized.
  \item Perform pop operation on $n_{in}$ and $n_{out}$ queues. Each round derives two nodes a and b, and create the edge $<a, b>$ thereby. Repeat such operation until the $n_{in}$ and $n_{out}$ queues are empty. At this point, the simulation is completed.
\end{enumerate}

\begin{figure}[htbp]
\begin{center}
\includegraphics[scale=0.5, angle=0]{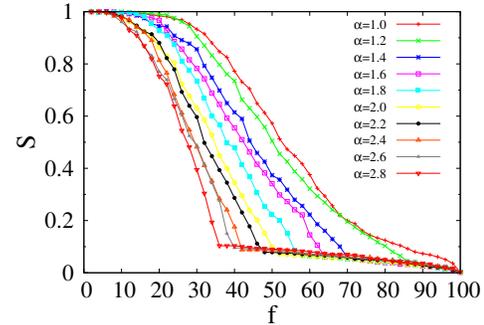}
\caption{Results of EA attack strategy in the generated networks with increasingly asymmetry. We obtained 10 random graphs each with about 800 nodes and 16,000 edges. The asymmetry of graphs grew as we increased $\alpha$ by a step of 0.2 from 1.0 to 2.8. The X axis is the percentage of the removed edges in the networks. The Y axis represents the attack effect.} \label{fig:6}
\end{center}
\end{figure}

In the above-mentioned algorithm, we add a parameter $\alpha$ to control the relative sizes of out-degree range and in-degree range so that $in_{min}=\alpha \times out_{min}$, $in_{max}=\alpha \times out_{max}$. When $\alpha$ increases from the initial value of 1, the in-degree range of network node will become larger than the out-degree range. The asymmetry of generation graph is adjusted accordingly. Figure \ref{fig:6} shows that, as the coefficient $\alpha$ increases, EA attack strategy achieves better results as asymmetry of the same scale networks becomes more prominent.


\begin{figure}[htbp]
\begin{center}
\subfigure[email-euall]{\includegraphics[width=.23\textwidth]{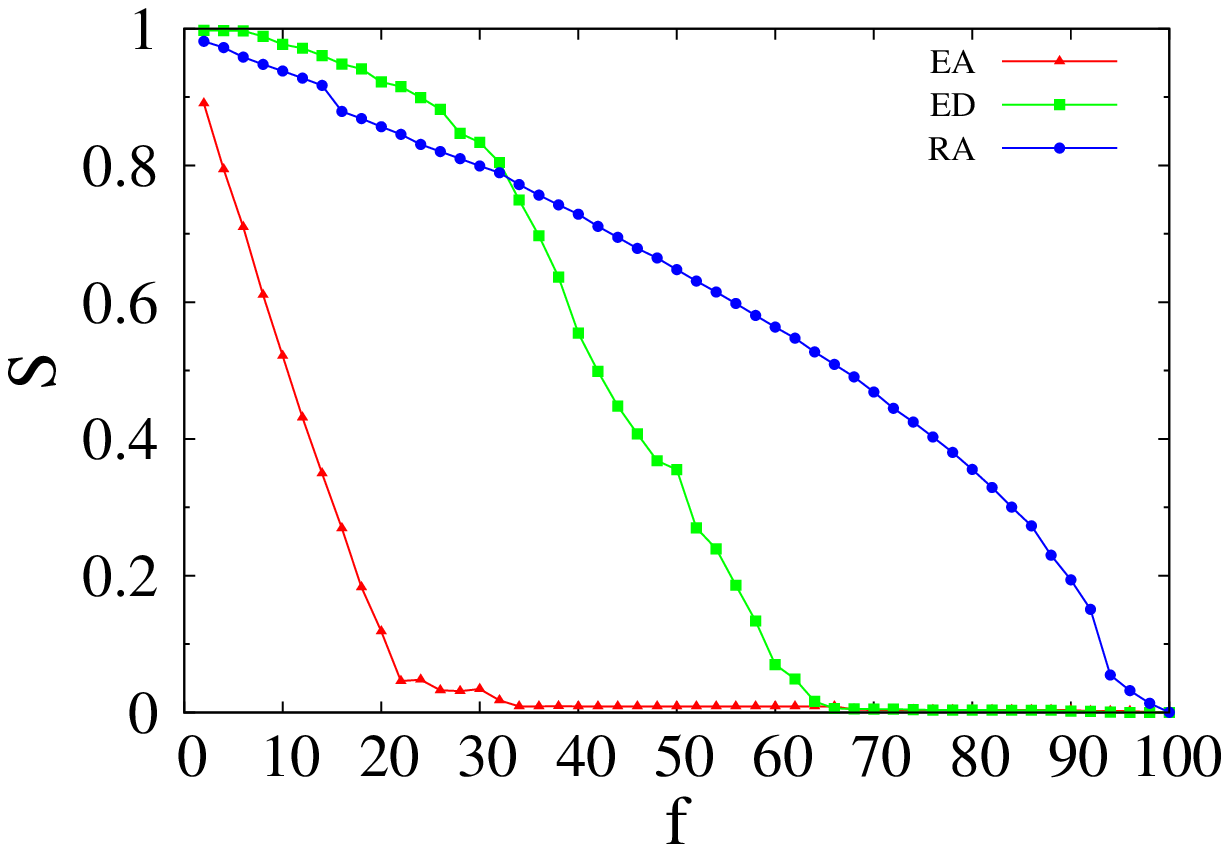}}
\subfigure[web-berkstan]{\includegraphics[width=.23\textwidth]{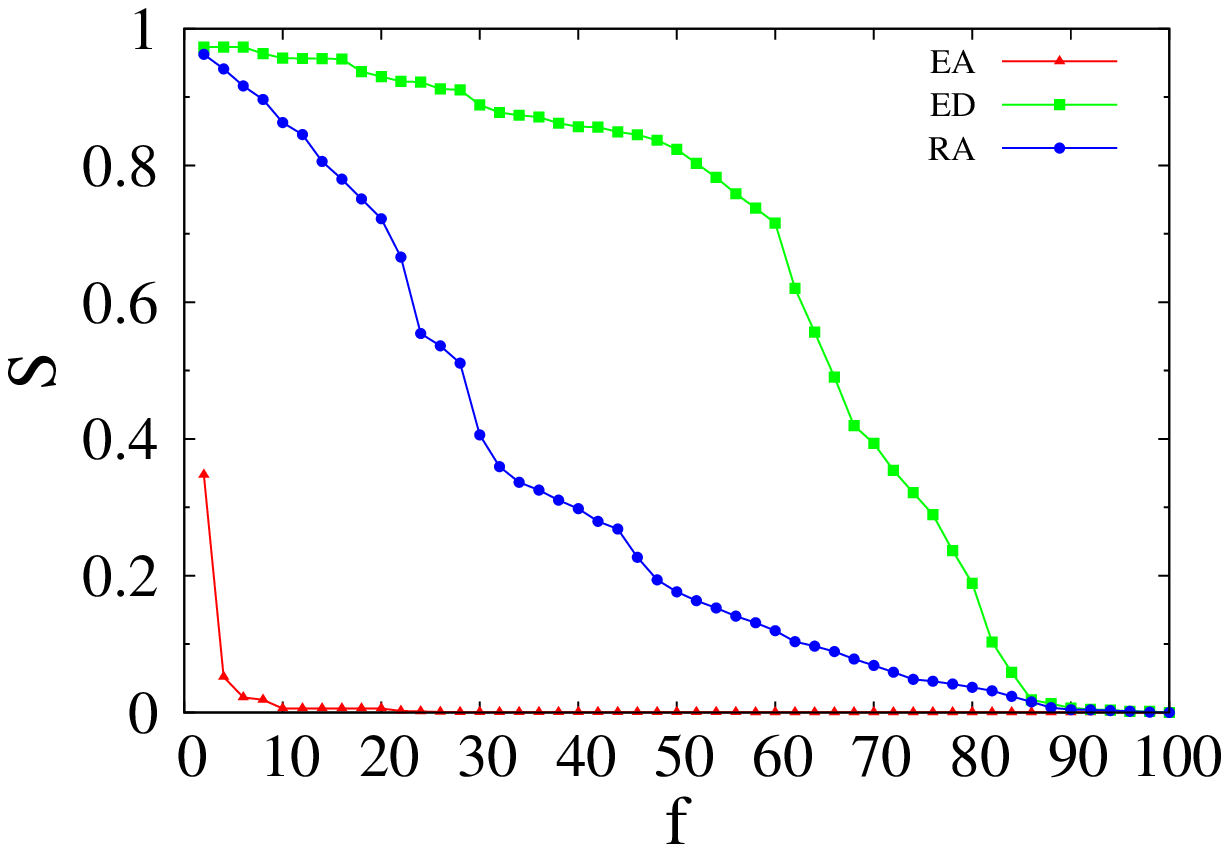}}
\subfigure[amazon0601]{\includegraphics[width=.23\textwidth]{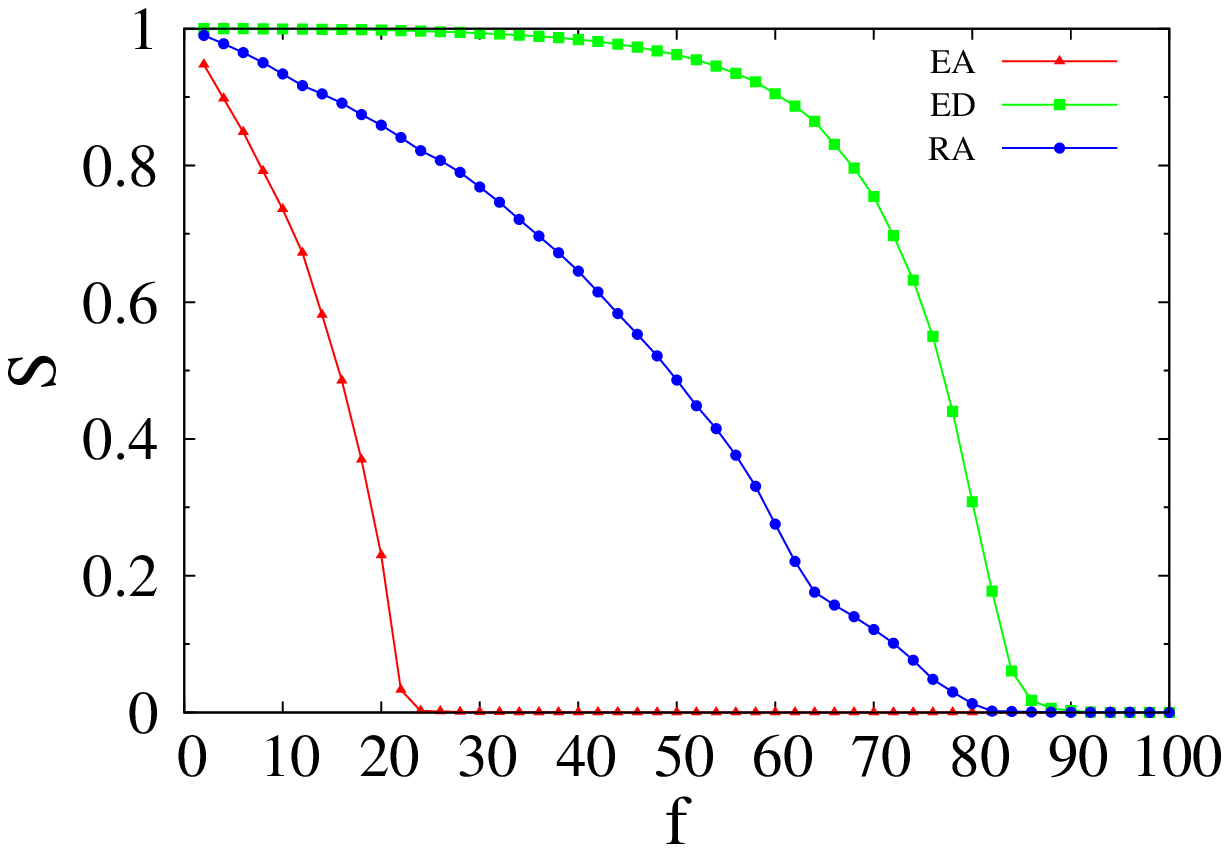}}
\subfigure[wikitalk]{\includegraphics[width=.23\textwidth]{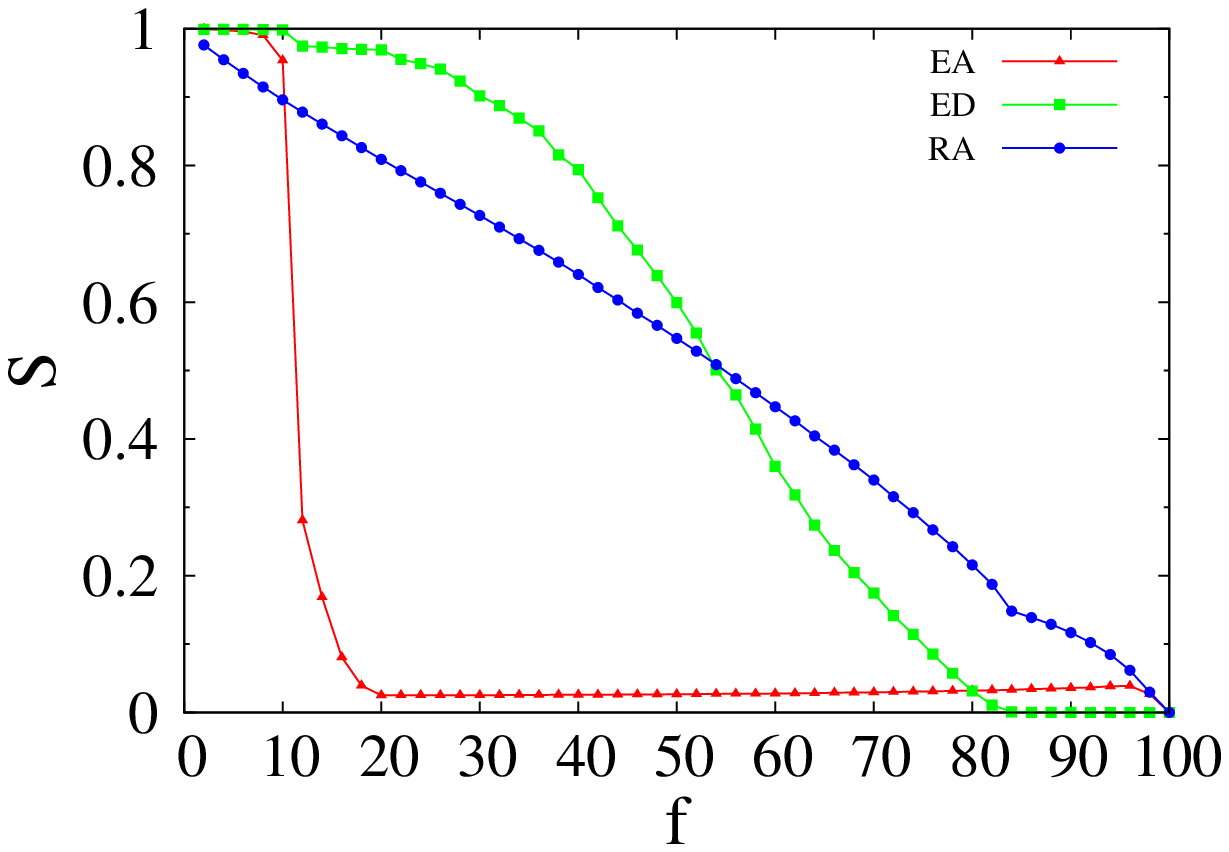}}
\caption{Results of attacks on large data sets. The data sets were Email-Euall\cite{Leskovec2007Graph} with 265,214 nodes and 420,045 edges, web-BerkStan\cite{Jure2009Community} with 685,230 nodes and 7,600,595 edges, amazon0601\cite{Leskovec2007The} with  403,394 nodes and 3,387,388 edges, and Wiki-Talk\cite{Leskovec2010Predicting} with 2,394,385 nodes and 5,021,410 edges. The X axis is the percentage of the removed edges in the networks, and the Y axis represents the attack effect.} \label{fig:7}
\end{center}
\end{figure}

In computational complexity, the EA attack strategy has a great advantage compared with the EB attack strategy. Computing edge betweenness is a critical step in the analysis of complex network, but its cost is very high. Generally, for the network with \emph{n} nodes, space complexity of computing edge betweenness is $O(n^{2})$ , and time complexity is $O(n^{3})$. We adopted the algorithm in Ref. \cite{Ulrik2001A} to compute edge betweenness. For a graph with \emph{n} nodes and \emph{m} edge, the space complexities of both weighted graph and unweighted graph are $O(m+n)$, the time complexity of weighted graph is $O(mn+n^{2}logn)$, and the time complexity of unweighted graph is $O(mn)$. In contrast, EA of each edge can be derived through the degree of nodes on both sides of the edge. Therefore EA can be obtained by traversing the edges. The time complexity of computing EA is $O(n)$ and the space complexity is $O(m)$ for a graph with \emph{n} nodes and \emph{m} edges. Thus, when attack effects are similar, the EA attack strategy is superior to the EB attack strategy in terms of computing cost. We selected four groups of large-scale data sets, including Email-Euall\cite{Leskovec2007Graph}, web-BerkStan\cite{Jure2009Community}, amazon0601\cite{Leskovec2007The} and Wiki-Talk\cite{Leskovec2010Predicting}, to conduct attack experiments. Because the amount of computation of edge betweenness-based attack strategy was too large, we only performed the other three attack strategies, i.e. EA, ED and RA strategies.



Figure \ref{fig:7} shows the results of attacks on large data sets. The EA attack strategy performed far better than the ED and RA attack strategies. The experiment setup consisted of Ubuntu14.04 32-bit system, GCC 4.8.4, and i7 processor with 2GB memory and 3.4GHz quad-core. The EA attack strategy took no more than 190 seconds to compute the largest-scale wiki-Talk data sets.



This work was supported by National Natural Science Foundation of
China (No. 61672073 and No. 61272167).

\bibliographystyle{apsrev4-1}
\bibliography{my}

\end{document}